# Performance of Modeling wireless networks in realistic environment

**Mohammad Siraj**  siraj@ksu.edu.sa
*Department of Computer Engineering*
*College of Computer and Information Sciences)*
*Riyadh-11543,SA*

**Soumen Kanrar**  soumen_kanrar@yahoo.co.in
*Department of Computer Engineering*
*College of Computer and Information Sciences)*
*Riyadh-11543,SA*

**Abstract:**

A wireless network is realized by mobile devices which communicate over radio channels. Since, experiments of real life problem with real devices are very difficult, simulation is used very often. Among many other important properties that have to be defined for simulative experiments, the mobility model and the radio propagation model have to be selected carefully. Both have strong impact on the performance of mobile wireless networks, e.g., the performance of routing protocols varies with these models. There are many mobility and radio propagation models proposed in literature. Each of them was developed with different objectives and is not suited for every physical scenario. The radio propagation models used in common wireless network simulators, in general researcher consider simple radio propagation models and neglect obstacles in the propagation environment. In this paper, we study the performance of wireless networks simulation by consider different Radio propagation models with considering obstacles in the propagation environment. In this paper we analyzed the performance of wireless networks by OPNET Modeler .In this paper we quantify the parameters such as throughput, packet received   attenuation.

## 1. Introduction:

Wireless communication technologies are undergoing very rapid advancements. In the past few years researcher have experienced a steep growth in the area of wireless networks in wireless domain. The attractiveness of wireless networks, in general, is attributed to their characteristics/features such as ability for infrastructure-less setup, minimal or no reliance on network planning and the ability of the nodes to self-organize and self-configure without the involvement of a centralized network manager, router, access point or a switch. These features help to set up a network fast in situations where there is no existing network setup or in times when setting up a fixed infrastructure network is considered infeasible, for example, in times of emergency or during relief operations. Even though wireless networks have emerged to be attractive and they hold great promises for our future, there are several challenges that need to be addressed. Some of the well-known challenges are attributed to issues relating to scalability, quality-of-service, energy efficiency and security. Currently the field of wirelesses communication systems is one of the fastest growing segments of the communications industry. Wireless communication systems, such as cellular, cordless and satellite phones as well as wireless networks  have found widespread use and have become an essential tool in  everyday life both professional and personal. Radio channels are more complicated to model than wired channels. Their characteristics may change rapidly and

Mohammad Siraj & Soumen Kanrar

randomly and they are dependent on their surroundings. The radio propagation models used in common wireless networks simulator is a free line of sight communication between the different devices assuming obstacle free area. As a result this range is nothing but a circle assuming that the devices residing within this circle receive the transmitted frames without errors. It is assumed that radio signals are completely blocked by obstacles. This approach poorly reflects radio wave propagation in a typical outdoor scenario, like a city center in which buildings significantly affect communication between nodes. Most of the publications investigating wireless networks behavior consider simple models. There exists a lot of work using NS2 but very few in OPNET platform [6] . A limitation in NS2 regarding two ray ground is that sender and receiver have to be in the same height [2,3,4,5]. The earlier work was carried out in open space with random mobility and idealized signal propagation model. In modeling authors investigate the behavior by randomly placing building environments. Secondly a scenario consists of several zones. Zones are either movement zones or obstacles[2,3,4]. Such environments are modeled not accurately. In this paper we have used CADRG/CIB (Compressed Arc Digitized Raster Graphics / Controlled Image Base) maps and terrain modeling module to achieve the realistic results by creating parameters sets of frequently used combinations of parameter values such as open terrain in dry weather, open terrain in rainy season and terrain with heavy vegetation by varying surface refractivity, relative permittivity, ground conductivity and resolution. Using opnet TMM (Terrain modeling module) we have simulated the radio signals being distributed over the varying terrain and observed the changes in signal strength. The signal strength is observed as the receiver moves over pre-defined path that goes through many elevation changes. TMM enhances the accuracy of the result by taking into account signal loss due to the effect of the terrain. Incorporating TMM we can determine whether the sender will be able to communicate with the receiver due to the terrain. It gives more accurate Propagation loss, signal strength and noise results.

In this paper we have shown the performance of wireless networks by integrating realistic models. We also observed how the models affect the output (i.e. Simulation results). In this paper we use OPNET Modeler 14.5. Here we integrate more realistic radio propagation models. Besides the model itself, we consider the geography data of the simulation area. In our case we have taken it from digital map vendors. The ionosphere affects radio signals in different ways depending on their frequencies (Figure 1), which range from extremely low (ELF) to extremely high (EHF). On frequencies below about 30 MHz the ionosphere may act as an efficient reflector, allowing radio communication to distances of many thousands of kilometers. Radio signals on frequencies above 30 MHz usually penetrate the ionosphere and, therefore are useful for ground-to-space communications. The ionosphere occasionally becomes disturbed as it reacts to certain types of solar activity. Solar flares are an example; these disturbances can affect radio communication in all latitudes. Frequencies between 2 MHz and 30 MHz are adversely affected by increased absorption, whereas on higher frequencies (e.g., 30–100 MHz) unexpected radio reflections can result in radio interference. Scattering of radio power by ionospheric irregularities produces fluctuating signals (scintillation), and propagation may take unexpected paths. TV and FM (on VHF) radio stations are affected little by solar activity, whereas HF ground-to-air, ship-to-shore, Voice of America, Radio Free Europe, and amateur radio are affected frequently. Figure 2 illustrates various ionospheric radio wave propagation effects.

Some satellite systems, which employ linear polarization on frequencies up to 1 GHz, are affected by Faraday rotation of the plane of polarization.

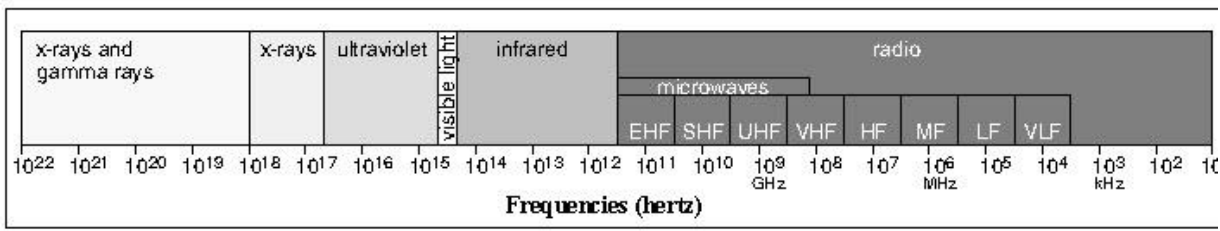

**Figure 1** : The Electromagnatic Spectrum

Mohammad Siraj & Soumen Kanrar

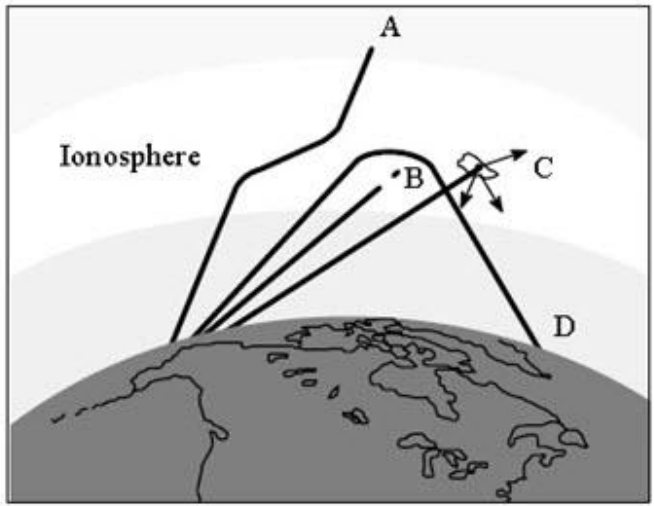

A. *Wave penetrates the ionospheric layer.*
B. *Wave is absorbed by the layer.*
C. *Wave is scattered in random directions by irregularities in the layer.*
D. *Wave is reflected normally by the layer.*

**Figure 2**: Radio Waves In Ionosphere

## 2 Preliminaries

**Free Space Model:** The Free Space model represents by equation (1) a signal propagating through open space, with no environmental effects. It has one parameter, called "lineofsight". With this parameter off, terrain has no effect on propagation. With it on, the model uses terrain data solely to determine if a line-of-sight (LOS) exists between the transmitting and receiving antennas. If there is no LOS, the signal is blocked entirely and no communication takes place.

$$P_r = \frac{P_t G_t G_r \lambda^2}{(4\pi)^2 d^2 L}$$ --------------------------------- (1)

Where $P_r$ the received signal is power (in Watt), $P_t$ is the transmitted signal power, $G_r$ and $G_t$ are the gains of the receiving and the transmitting antennas respectively. $\lambda$ is the wave length, L is the system loss, and d is the distance between the transmitter and the receiver. According to [8], a single direct path between the communicating partners exists seldom at larger distances

**Two Ray Ground Model:** The two ground model equation (2) assumes that the received energy is the sum of the direct line of sight path and the path including one reflection on the ground between the sender and the receiver. The received power becomes independent of the frequency of the transmitted signal but depends on the height of the transmitter $h_t$ and the receiver $h_r$, $G_r$ and $G_t$ are the gains of the receiving and the transmitting antennas respectively, d is the distance between the transmitter and the receiver.

$$P_r = \frac{P_t G_t G_r h_t h_r}{d^4}$$ -------------------------------------- (2)

### 2.1 Total Path loss:
The total path loss $L_t$ in db is defined as

$$L_t = L_o + L_l + L_{md} + L_r$$ --------------------------------- (3)

Mohammad Siraj & Soumen Kanrar

$L_o$, $L_l$ and $L_{md}$ are the losses due to free space propagation, local screen ( building) in the vicinity of the mobile node and multiple diffractions caused by the building respectively. $L_r$ is the loss caused by the reflection of the diffracted electric fields from the walls of the building next to mobile.

## 2.2 Line of sight (Los) propagation model:

When the mobile node is located in the vehicle over the flat terrain, Los propagation using 2-ray model .The received power can be expressed as-

$$\frac{P_r}{P_t} = \left\{\frac{\lambda}{4\pi}\right\}^2 \left| \frac{1}{r_1}\exp(-jkr_1) + R\frac{1}{r_2}\exp(-jkr_2) \right|^2 \text{------------------ (4)}$$

Where $P_t$ and $P_r$ denote the base station or transmitter power and power received by the receiver mobile node. R is the reflection coefficient and $r_1$, $r_2$ represents the path length of the direct ray and the path length due to ground reflected ray, k is constant.

## 2.3 Troposcatter loss model:

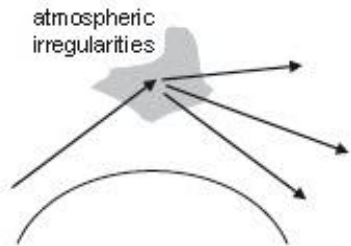

**Figure 3:** Troposphere Scattering

Air is not uniform, there are eddies, thermals, turbulence etc. where the air has slightly different pressure and hence a different refractive index. Scattering from refractive index irregularities in the high atmosphere (the troposphere) with sufficiently directive antennas and high transmitter powers is called *troposcattering* shown by figure-3

The median loss is given by
L = M + 30log(f) + 10log(d) + 30log(θ) + LN+ LC - Gt – Gr -----------(5)
This is an empirical model, with M is typically 19-40 dB depending on climate.
Where θ is the scatter angle (milliradians), LN is the height of the common volume, LC is the aperture-medium coupling loss and Gt , Gr are the gains of the antennas.

## 2.4 Diffraction loss:

If the direct line –of-sight is obstructed by a single knife-edge type of obstacle, with height $h_m$ via –figure Where $T_x$ and $R_x$ are the transmitter and receiver at the points B and C. |BE| = $d_T$,

|CE|= $d_R$ , |DE|= $h_m$

Mohammad Siraj & Soumen Kanrar

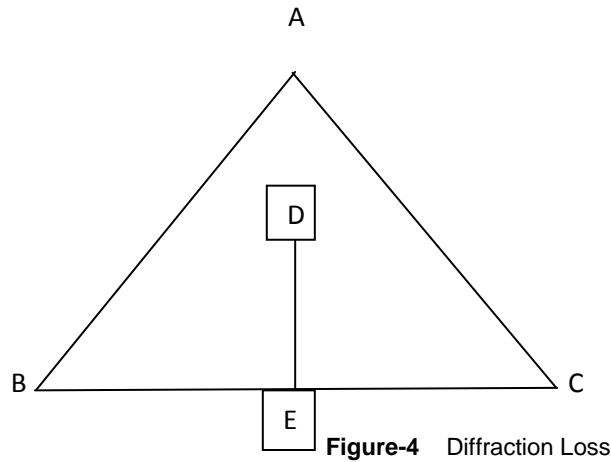

**Figure-4** Diffraction Loss

diffraction parameter $v$ can be expressed as,

$$v = h_m \left\{ \frac{2}{\lambda} \sqrt{(\frac{1}{d_T} + \frac{1}{d_R})} \right\} \quad \text{----------------------------------(6)}$$

The diffraction loss additional to free space loss and express in dB can be closely approximated by

$A_D = 0$   if $v < 0$

$A_D = 6 + 9v + 1.27v^2$   if $0 < v < 2.4$

$A_D = 13 + 20 \log v$   if $v > 2.4$

The attenuation over rounded obstacles higher than $A_D$ in the above formula

## 2.5 Rician Fading Model :

Rician fading is a stochastic model for radio propagation anomaly caused by partial cancellation of a radio signal by itself. The signal arrives at the receiver by two different paths( hence exhibiting multipath interference), and at least one of the paths is changing( lengthening or shortening). Rician fading occurs when one of the paths, typically a line of sight signal is much stronger than the others. In Rician fading ( via fig -3)a strong dominant component is present. Similar to the case of Rayleigh fading, the in-phase and quadrature phase component of the received signal are i.i.d. jointly Gaussian random variables. However, in Rician fading the mean value of (at least) one component is non-zero due to a deterministic strong wave.

Mohammad Siraj & Soumen Kanrar

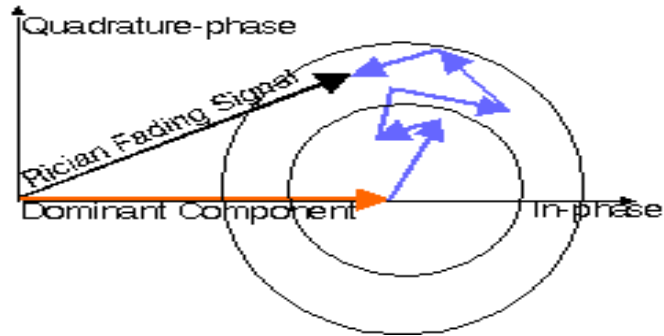

**Figure 3**: Rician Fading Model

The amplitude gain at the receiver is characterized by a Rician distribution.

The Rician amplitude r with parameter K can be defined as

$$r = \sqrt{(\sigma x_1 + A)^2 + \sigma x_2^2} \quad \text{------------------------------- (7)}$$

Where $x_1$ and $x_2$ are quadrature components and $A^2/(2\sigma^2) \cong K$, $\sigma x_i$ is a random variable with variance $\sigma^2$. The mean-squared value of the Rician distribution is given by

$$A^2 + 2\sigma^2 = 2\sigma^2(K+1) \quad \text{------------------------------- (8)}$$

and we get the normalized power envelop from (7) and (8) is

$$\frac{r^2}{P} = \frac{1}{2(K+1)}\left[(x_1 + \sqrt{2K})^2 + x_2^2\right] \quad \text{----------------------(9)}$$

Where P is the large scale model expression for the power in the dominant path. This power envelop can be used to modulate the output of a large scale propagation model. The above description assumes that the mean squared value of the envelop is the power predicted by the large scale model. Specifically, in the Rician model, this power contains the power in the dominant path and the multipath power. With certain propagation models, it may be more appropriate for the large scale power calculation to represent only the value in the dominant path. In the equation (7) after divided by the factor $A^2 = 2\sigma^2 K$. The normalized power envelop then takes the form

$$\frac{r^2}{P} = \frac{1}{2K}\left[(x_1 + \sqrt{2K})^2 + x_2^2\right] \quad \text{-----------------------------(10)}$$

Where P is the large scale model expression for the power in the dominant path.

### 2.6 Longley Rice Model :

The Longley-Rice model predicts long-term median transmission loss over irregular terrain relative to free-space transmission loss. The model was designed for frequencies between 20 MHz and 40 GHz and for path lengths between 1 km and 2000 km. This model considers environmental condition along the signal path. Here we can create parameters sets to hold frequently used combination of parameters values such as .

|  | Relative Permittivity | Conductivity (Siemens per meter) |
|---|---|---|
| Average ground | 15 | 0.005 |
| Poor ground | 4 | 0.001 |
| Good ground | 25 | 0.020 |
| Fresh water | 81 | 0.010 |
| Sea water | 81 | 5.000 |

**TABLE 1** : Longley Rice Parameters

### 2.7 TIREM :

The Terrain-Integrated Rough Earth Model is a computer software library that calculates basic median propagation loss(path loss) of radio wave over irregular earth terrain. The method was develop revived improved and evolved into a TIREM software version for use by the Department of Defense (DoD). The TIREM used for radio frequencies in the range of 1 through 20,000 MHz over terrain elevations which are specified by a set of discrete points between the great circle path of the transmitting antenna and receiving antenna. The earth terrain information can be provided by the digital terrain elevation data(DTED). TIREM provides more accuracy in the radio propagation model than the FSPL model by taking into account the transmitting medium (surface refractivity and humidity), antenna properties (height, frequency, and polarization), conductivity, and terrain elevations). The calculation of path loss is also determined by effects of free space spreading, reflection, surface wave propagation, diffraction, tropospheric scatter propagation, and atmospheric absorption but not ducting phenomena, fading, ionospheric effect, or absorption due to rain or foliage.

| Variable | Description | Valid Range |
|---|---|---|
| CONDUC | Conductivity of earth surface | 0.00001 to 100 S/m |
| EXTNSN | Profile indicator flag : .TRUE –profile is an extension of the last path along a radial .FALSE -new profile | .TRUE or .FALSE |
| HPRFL | Array of profile terrain heights above mean sea level. | -450 to 9000 m |
| HUMID | Surface humidity at the transmitter site | 0 to 50 g/m$^3$ |
| NPRFL | Total number of profile points for the entire path | $\geq 3$ |

Mohammad Siraj & Soumen Kanrar

| | | |
|---|---|---|
| PERMIT | Relative Permittivity | 1 to 100 |
| POLARZ | Transmitter antenna polarization : 'V' –Vertical 'H'-Horizontal | 'V' or 'H' |
| PROPFQ | Transmitter frequency | 1 to 20000MHz |
| RANTHT | Receiver structural antenna height | >0 to 30000m |
| REFRAC | Surface refractivity | 200 to 450 N-units |
| TANTHT | Transmitter structural antenna height | >0 to 30000 m |
| XPRFL | Array of great-circle distance from the beginning of the profile Point to each profile point | $\geq$ 0 m |

**TABLE 2:** TIREM Parameters Ranges

To develop the Algorithm for the TIREM we consider the equations (3) ,(4),(5) and (6)

-------------------------------------------------------------------------------------------------------------------------------

**Algorithm for the TIREM**

    Step-1 :  Input the parameters from the set of parameters

    Step-2:   Extract the key parameters

    Step-3:   If (line of sight) == True

    Step-4:   compute line of sight loss

    Step-5:   go to step- 10

    Step-6:   If (line of sight) == False

    Step-7:   compute diffraction loss

    Step -8:  compute Troposcatter loss

    Step-9:  combine diffraction loss and Troposcatter loss

    Step-10: compute mode and total path loss

    Step-11: output

-------------------------------------------------------------------------------------------------------------------------------

## 3. Simulation Model:

OPNET used to build the simulation model. All the operations are done by using OPNET kernel procedures. This is Baseline simulation. Till now we had been using default radio pipeline of OPNET figure - 4. The radio pipeline stages model the flow of data from a transmitter to a receiver. There are stages like transmission delay pipeline stage, SNR stage, BER stage etc. To realize the Rician fading stage 7 (receiver power ) of the Radio Pipeline stage *figure-4* has been modified. This stage uses by default free space propagation model. The purpose of this stage is to compute the receive power of the arriving packet's signal (watts). For packets that are classified is valid, the received power result is the key factor in determining the ability of the receiver to correctly to capture the information in the packets .Packets are classified as valid , invalid , or ignored by the stage -3 via-figure-4. In general, the calculation of received power is based on factors such as the power of the transmitter , the distances separating the transmitter and the receiver , the transmission frequency ,and transmitter and receiver antenna gain. The simulation kernel requires that the received power stage procedure accept a packet address as its sole argument. The antenna pattern and the power parameters have been defined at the runtime of the simulation.

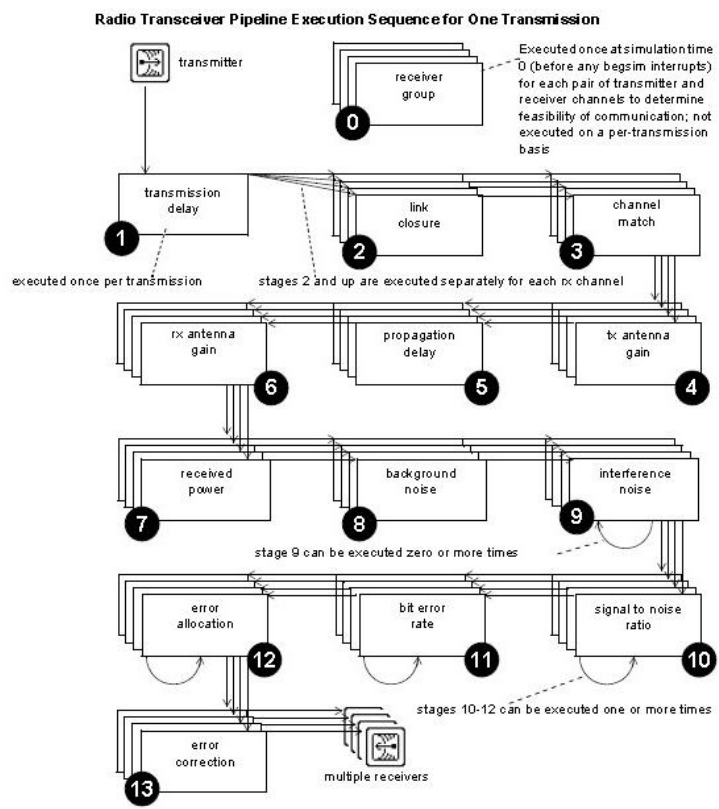

**FIGURE 4:** Radio Pipeline Stages

So for realizing the Rician fading we used the Rician Fading Power Module as via figure 4.

### 3.1 Node model

**3.1.1 Transmitter :** The transmitter composed of three modules via figure -5

1) Simple source i.e. packet generator.
2) Radio transmitter module. This module transmits the packets to the antenna at 1024 bits/sec using 100 percent of its channel bandwidth .

3) Antenna module. This module transmits signals.

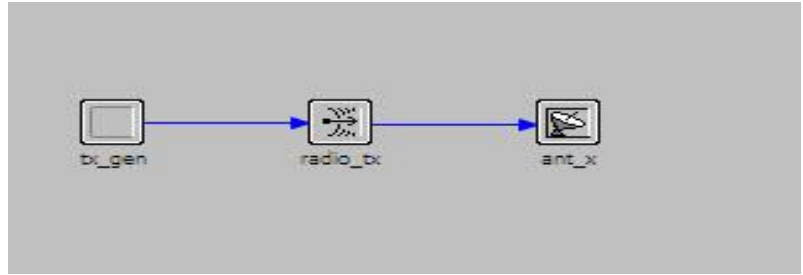

**FIGURE – 5** Transmitter

Fig: 6 represent Transmitter ($T_x$) Node Model. The parameters which are shown below in the table -2 have been used for Transmitter which is shown in figure – 6

| Packet Format | None |
|---|---|
| Packet Inter arrival Time | constant1.0 |
| Packet Size | Constant 1024 |
| Start Time | 10 |

**TABLE -3** : Simulation parameters for Transmitter

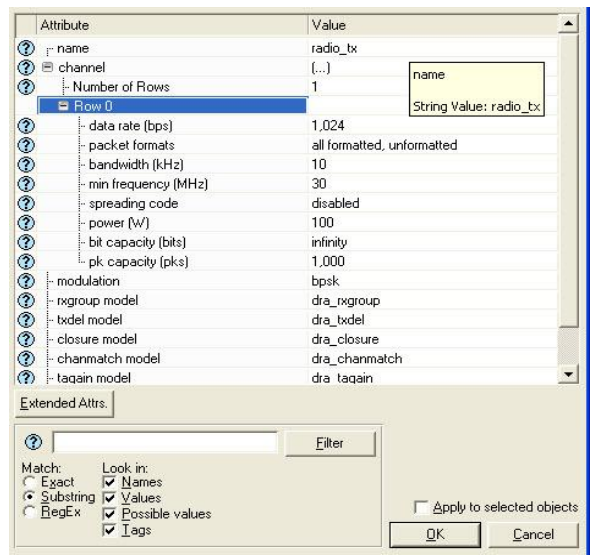

**FIGURE 6:** Transmitter $T_x$ attributes

This $T_x$ module is used for the all propagation models.

### 3.1.2 Receiver:

The receiver composed of three modules via figure- 7

1) Antenna module.
2) Radio receiver module: The radio receiver module consider several properties for each arrival of packets to determine, if the packet's average bit error rate (BER) is less than a specified threshold. If the BER is low enough the packet is sent to the sink and destroyed. This module defines the gain which will be adjusted at run time.
3) Sink processor module: This module store the received packets.
4) Processor module ( $R_x$ ) which calculates the information that the antenna needs to point    at a target :

    latitude, longitude and altitude coordinates the processor makes this calculation by using a kernel procedure that convert a node position in a subnet ( described by the X and Y position attributes) into the global coordinates that the antenna required.

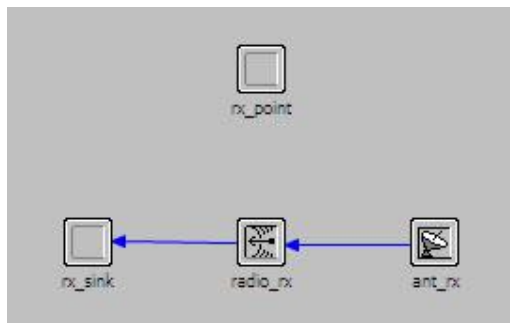

**FIGURE 7:** Receiver Antenna

The parameters which are shown below in the table - 2 have been used for the Receiver $R_x$ for Free Space Model, Hata Large City, Longley Rice and Risen Model

| Network Scale | Enterprise |
|---|---|
| Size | 8 X 5 Km |
| Packet Interarrival Time | Constant 1.0 |
| Packet Size | Constant 1024 |
| Start Time | 10 |
| Data rate(bps) | 1,000,000 |
| Bandwidth(KHz) | 20,000 |
| Minimum Frequency(MHz) | 905 |
| Spreading Code | Disabled |
| Modulation | bpsk |
| Power | promoted |

**TABLE 2:** Simulation Parameters For Receiver

Figure-8 is the $R_x$ attributes used in for Free Space Model, Hata Large City and Longley Rice models

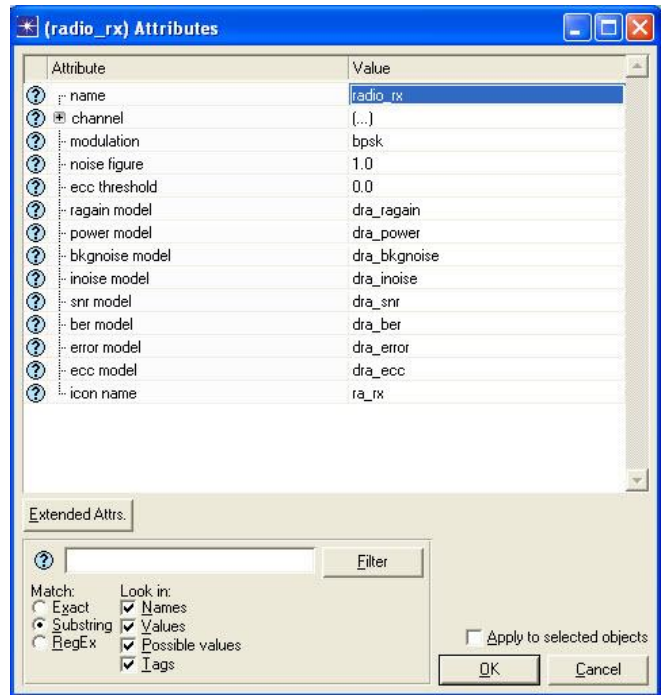

**Figure 8:** $R_x$ Attributes

Figure-9 is the $R_x$ attributes used in Rician models. The parameters which are shown below in the table - 3 have been used for the Receiver $R_x$ for Rician Model

| Attribute | Value for Rx |
|---|---|
| Altitude | 0.003 |
| Radio .rx Max Velocity | 1.0 |
| Radio rx Table Offset | 0 |
| Radio_ rx. Rician K Factor | 0.5 |
| Radio _rx .Use Two Ray | 1 |

**TABLE  3:**   Simulation Parameters For Risen Propagation Model

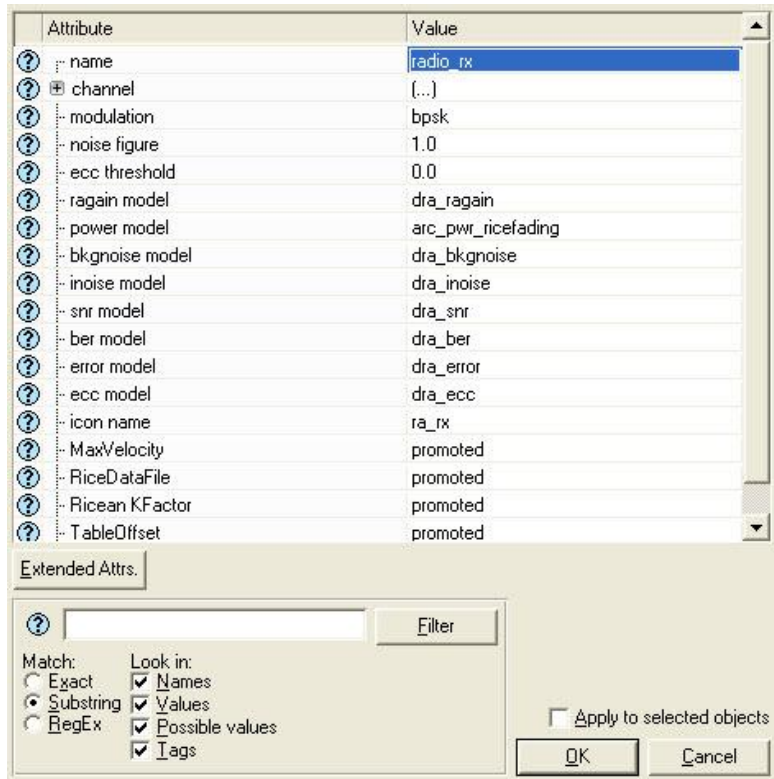

**FIGURE 9:** Attributes For Rician Propagation Model

## 4. Results and Discussion:

The following graphs via fig -10.1, 10.2, and 10.3 represents the Attenuation vs Distance. The graph are corresponding to the different propagation models free space ,Longley Rice model and TIREM model respectively. The transmitter and receiver are vertically polarized at the height of 25m (fig-10.1). We see further that the receiver, from the transmitter attenuation increases. From the fig-10.1 we observe the free space model shows linearity between the distance and the attenuation. Since the Longley Rice and TIREM takes terrain effect into consideration attenuation is much higher and realistic.

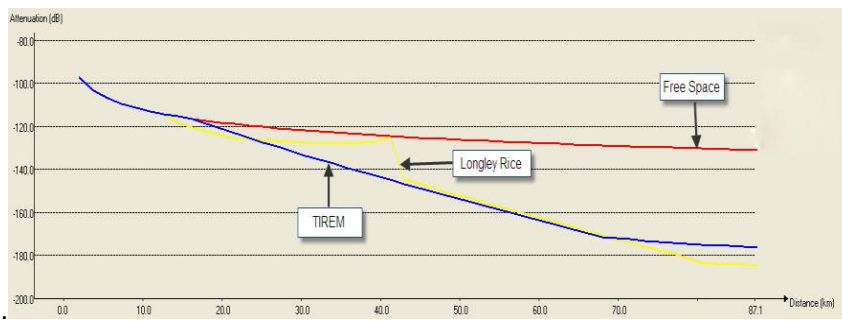

**FIGURE -10.1** Free space (Attenuation Vs Distance)

Antenna Height – 25 M

Mohammad Siraj & Soumen Kanrar

In the fig-10.2 and 10.3 the transmitter antenna height was raised to 75m and 100m respectively. By raising the height shows a consideriable change in signal strength .The attenuation in Longley Rice model is almost same as the free space model till 60Km (approx) after which it start separating and is same as TIREM model. So we see an elevation change affects signal strength.

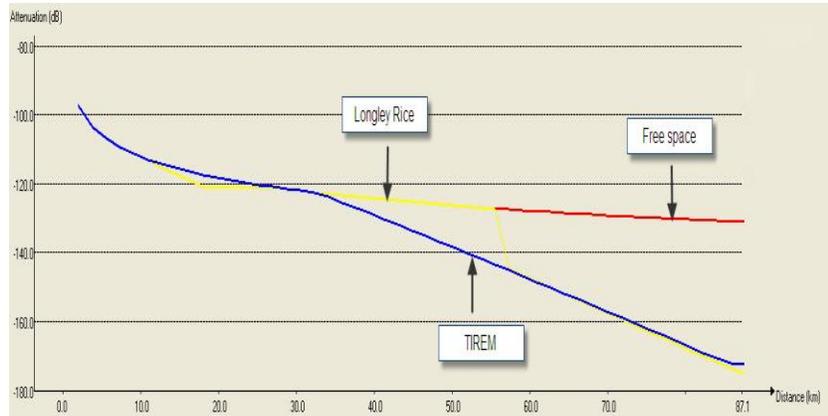

**FIGURE-10.2** Longley Rice (Attenuation Vs Distance)

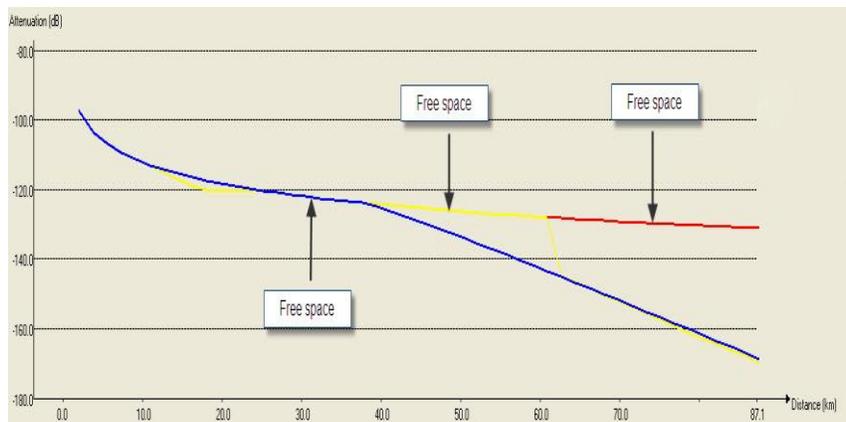

**FIGURE-10.3** TIREM    (Attenuation Vs Distance)

Figure -11 shows the traffic received at the receiver end .We observe diferent types of graph depending on the propogation model used.In the free space we notice not much change in the packet received. This is because that model assumes ideal propogation enverioment. The other two models (Longley Rice and TIREM models) are realstic model so we see changes in the graph. We observe packets get drop because of lower height of transmitter antenna.

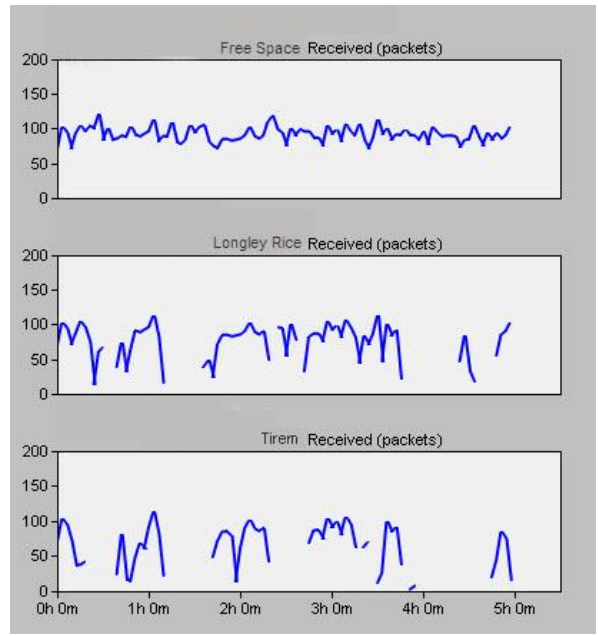

**FIGURE -11**   Packet  Received

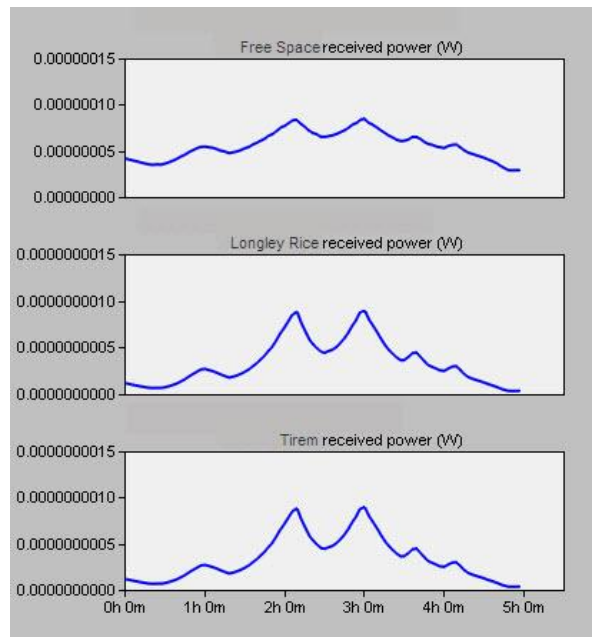

**Figure 12:** Received Power

The received power is only dependent on the transmitted power, the antenna's gain and on the distance between the sender and the receiver. It accounts manly for the facts that a radio signal which moves away from the sender has to cover in large area.So the received power decreasese with the square of the distance. Since the free sapce propogation model assume the ideal propogation condtion that there is only one clear line –of-sight path between the transmitter and the receiver the graph for the free space does not show much variation (Fig-12). In free space, the power radiated by an *isotropic antenna* is spread uniformly and without loss over the surface of a sphere surrounding the antenna. There is a considerable change in the received power for Longlay Rice and The TIREM models as it takes in to account the system loss (Fig-12). The path loss and the statistical characteristics of the received signal envelope are the most inportant for the coverage planing application.

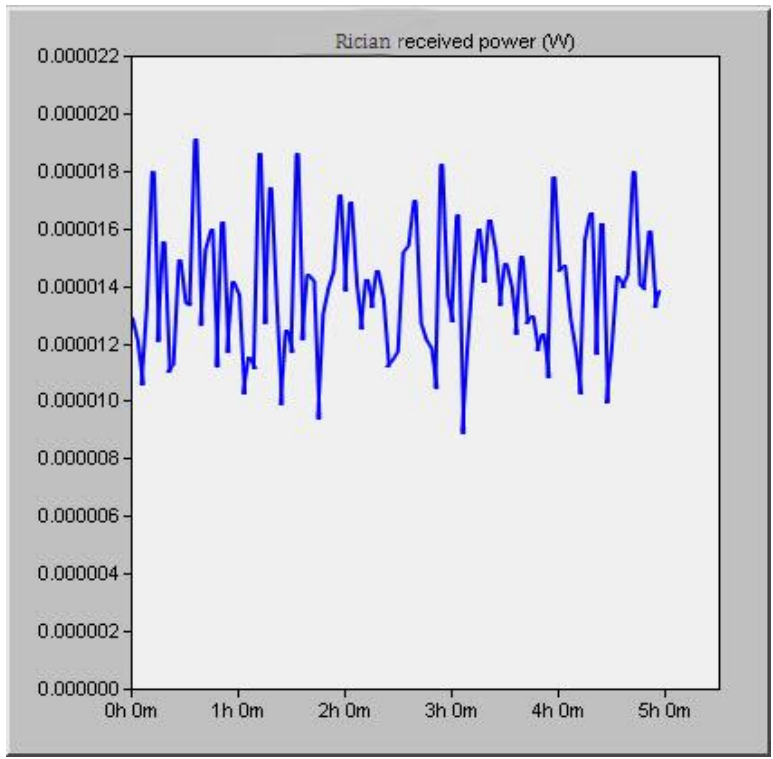

**FIGURE 13:** Power Level Of Received Packet

Figure-13 shows the power levels of the received packets of Rician fading model. Here we observe the realistic Rician fading power envelope at the receiver. This envelope is time correlated as its obeys the Risen distribution. Fading is mostly coused by the the multipath propogation of the radion signals. Rician fading occurs if there is one dominant( Line of Sight) path and multiple indirect signals . The above graph represent by the the expression

$$Y(t) = \sum_{n=0}^{N-1} \rho_n e^{j\varphi_n} \delta(t - \tau_n)$$ ---------------------------------(11)

Where N is the number of received signals (correspond to the number of electromagnetic paths and very large). $\tau_n$ is the time delay of the generic $n^{th}$ impulse and $\rho_n e^{j\varphi_n}$ represent the complex amplitude (i.e. magnitude and phase) of the generic received pulse. As a consequence $y(t)$ represents the impulse response of multipath correspond to the signal received by the receiver in Watt and x axis represent the time axis.

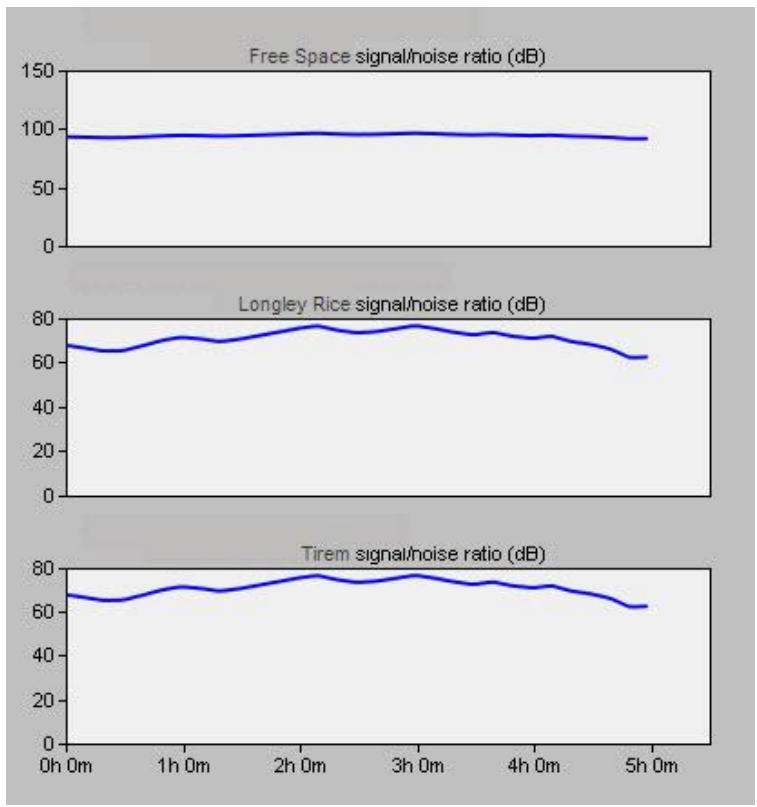

**FIGURE 14 :** Signal To Noise Ratio

The signal to noise ratio is important in the transmission of data because it set the upper bound on the achievable data rates .This is an important characteristic to show the performance of the wireless networks. There is a significant difference in the path loss between free space and the other two models ( fig 14). This difference is because Longley Rice and TIREM takes into account the real environment and the path loss calculation of each model. So with the change in terrain, SNR varies.

## 5.Concluding Remarks:

In this work we have shown the effect on the signal strength as the receiver moves over the varying terrain ( using *Compressed Arc Digitized Raster Graphics / Controlled Image* Base maps). This results give realistic result on propagation loss, signal strength and noise. We have also shown the effect on radio signal strength due to change in antenna height, elevation change and the distance between transmitter and receiver. In this work we observe Rician fading best characterizes a situation where there is a direct LOS path in addition to a number of indirect multiple singles. The Rician model is often applicable in an indoor environment. The Rician model also becomes more applicable in smaller cells or in more open outdoor environments. The channels can be characterized by a parameter K, can be express as the ratio of the (power in the dominant path) to the (power in the scattered paths). When K =0 the fading occurs when there are multiple indirect paths between transmitter and receiver and no distinct dominant path, such as an LOS path . Changing Max Velocity parameters will cause Rician fading envelope to fade faster or slower.

As further work more detailed simulation scenarios will be created by different performance evaluation by compare the radio propagation models, performance variety of metrics, Packets sent, throughput, dropped packets, packet Delivery Ratio and packet routing overhead. The more accurate movement and communication patterns could give

Mohammad Siraj & Soumen Kanrar

hints where existing protocols still have drawbacks and what has to be changed in order to overcome these problems. Especially when QoS – aware algorithms.